\documentclass[twocolumn,footinbib,showpacs,preprintnumbers,amsmath,amssymb,pra,superscriptaddress,longbibliography]{revtex4-1}
\usepackage{epsfig,subfigure,amssymb,amscd,amsmath,graphicx,amsfonts,mathtools,mathcomp,pbox}
\usepackage{times,longtable,math dots,hyperref,enumitem}
\usepackage{mathrsfs,physics}
\newcommand{\be}{\begin{equation}}
\newcommand{\ee}{\end{equation}}
\usepackage{tabularx}

\usepackage[usenames]{color}

\usepackage[usenames,dvipsnames]{xcolor}
\usepackage{soul}
\setulcolor{BrickRed}

\begin{document}

\title{Constrained thermalisation and topological superconductivity}

\author{S. Nulty}
\affiliation{ Department of Theoretical Physics, National University of Ireland, Maynooth, Co. Kildare, Ireland}
\author{J. Vala}
\affiliation{ Department of Theoretical Physics, National University of Ireland, Maynooth, Co. Kildare, Ireland}
\author{D. Meidan}
\affiliation{Department of Physics, Ben-Gurion University of the Negev, Beer-Sheva 84105, Israel}
\author{G. Kells}
\affiliation{ Dublin Institute for Advanced  Studies, School of Theoretical Physics, 10 Burlington Rd, Dublin, Ireland}
\date{\today}
\begin{abstract}
We examine the thermalisation/localization trade off in an interacting and disordered Kitaev model, specifically addressing whether signatures of many-body localization can coexist   with the systems topological phase.  Using methods applicable to finite size systems, (e.g. the generalized one-particle density matrix, eigenstate entanglement entropy,  inverse zero modes coherence length)  we identify  a regime of parameter space in the vicinity of the non-interacting limit where topological superconductivity survives together with a significant  violation of Eigenstate-Thermalisation-Hypothesis (ETH) at finite energy-densities. We further identify an  anomalous behaviour of the von Neumann entanglement entropy which can be attributed to the prethermalisation-like effects that occur due to lack of hybridization between high-energy eigenstates  reflecting   an approximate particle conservation law.  In this regime the system tends to thermalise to a generalised Gibbs ensemble (as opposed to the grand canonical ensemble).  Moderate disorder tends to drive the system towards stronger hybridization and a standard thermal ensemble, where the approximate conservation law is violated. This behaviour is cutoff by strong disorder which obstructs  many body effects thus  violating ETH and reducing the entanglement entropy. \end{abstract}


\maketitle

\section{Introduction}
The possibility of realising Majorana bound states in proximity coupled systems \cite{Fu2008,Lutchyn2010,Oreg2010,Duckheim2011,Chung2011,Choy2011,Kjaergaard2012,Martin2012,Nadj2013} has spurred a great deal of activity over the last decade \cite{Mourik2012,Deng2012,Das2012,Finck,Churchill2013,Albrecht,Gul2018,Deng2016,Nadj2014,Ruby2015,Pawlak2016}. Much of this stems from the belief that quantum devices based on  symmetry protected topological order (SPTO), are a promising platform for quantum information processing as they are robust against many common forms of decoherence \cite{Kitaev2001,Kitaev2003,Kitaev2006,Nayak2008,Stanescu2016}. 

With the aspiration  of enhancing the stability of these systems at higher temperatures, in recent years there has also been a growing interest in the behaviour of these systems at energies well above the topological gap. In this respect,  one exciting prospect is that it may be possible for regions of the topological parameter space to coincide with interacting regions that violate the Eigenstate Thermalisation Hypothesis (ETH) \cite{Deutsch1991,Srednicki1994,Tasaki1998} due to localisation in Fock-space {\cite{Bauer2013, Huse2013, Kells2018}. Where this can be engineered, it is expected that systematic errors due to non-adiabatic manipulation  (see \cite{Cheng2011,Scheurer2013,Perfetto2013,Bauer2018, Conlon2019})  and/or interaction with the environment  (see \cite{Goldstein2011,Budich2012, Rainis2012, Cheng2012,Schmidt2012, Konschelle2013, Mazza2013, Ho2014, Ng2015, Pedrocchi2015,Campbell2015,Huang2018,Knapp2018,Aseev2018, Aseev2019}) can be suppressed.   

The goal of this paper is to find out if/where the topological-order and many-body localising behaviour overlap in the interacting Kitaev chain (or Kitaev-Hubbard), the proto-typical model used to study this effect.  One challenging aspect here is that the key parameters being examined, namely disorder \cite{Motrunich2001,BDRvOProb2011,BDRvOTop2011,Akhmerov2011,RKDM2012,RBA2013,DeGottardi2013,Pientka2013} and interaction strength \cite{Stoudenmire2011,Sela2011,Lutchyn2011,Lobos2012,Crepin2014,Hassler2012,Thomale2013,Katsura2015,Gergs2016}, both eventually conspire to drive the system out of the topological phase. On the other hand, these effects work in opposition when considering localisation induced violations of ETH and hence, for a given interaction strength, one has to consider whether the disorder required to prevent thermalisation is weak enough to keep the topological properties of the phase intact \cite{Kells2018} .   

\begin{figure}
\centering
\includegraphics[width=0.35\textwidth]{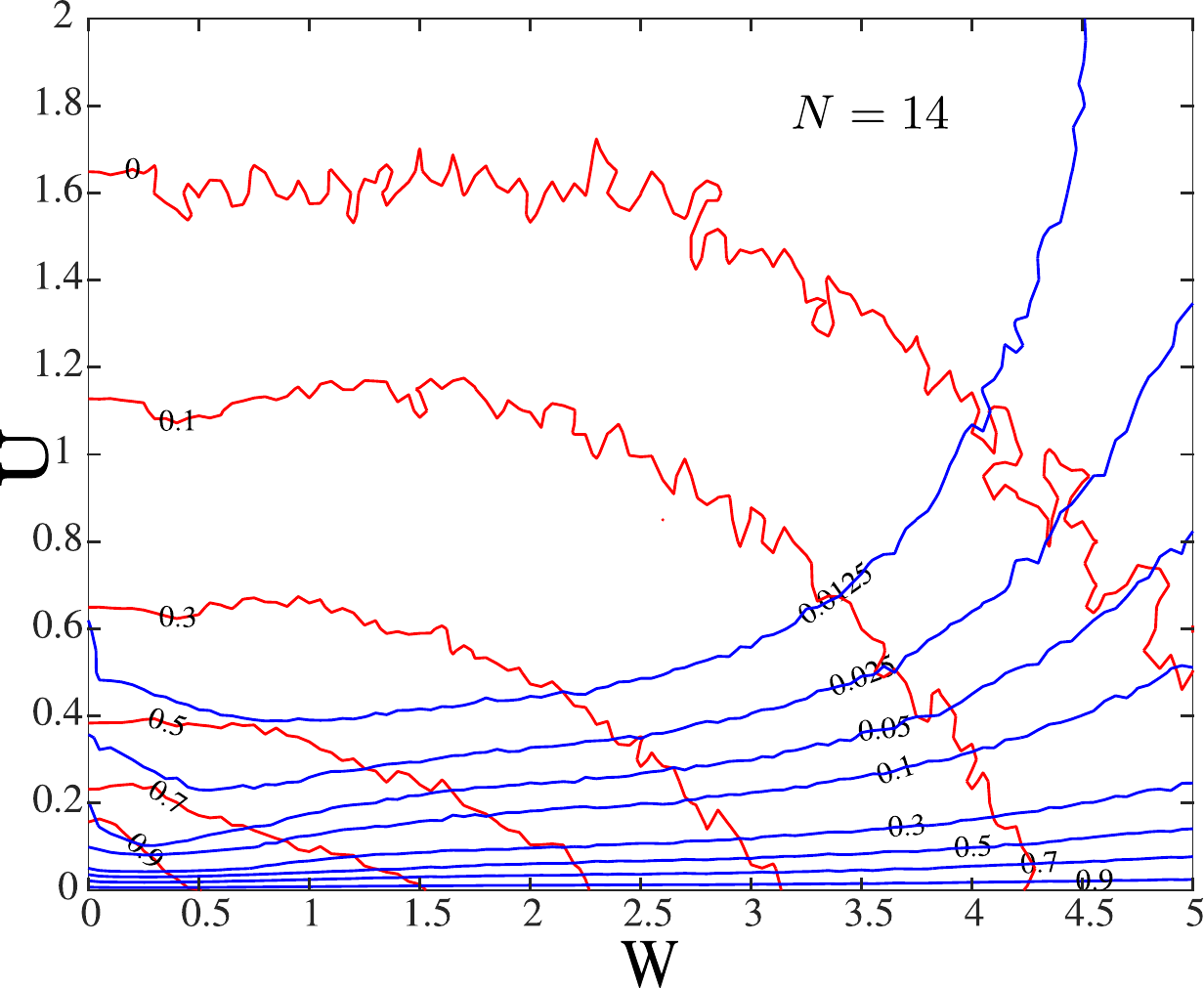}
\caption{ Average inverse coherence length $\langle 1/\xi\rangle$ (red) and occupation number discontinuity $\langle\Delta n\rangle$ (blue), plotting with increasing disorder strength $W$ and interaction strength $U$: system size $14$ sites, $\mu=0.4$, $\Delta=0.7$, $500$ disorder realizations. Due to the finite size of the system, one expects $\Delta n\sim O(1/N)$ to be indicative of the ergodic transition, while the edge of the topological region is in principle  indicated by values of $\langle 1/\xi\rangle  \to O(1/N)$. }
\label{fig:phases14}
\end{figure}

Our first result is that  there are indeed regions of parameter space where both MBL signatures and topological order can co-exist, although these regions can generally only occur in a narrow window about the non-interacting regime. The key result in this regard is the contour plot  Figure \ref{fig:phases14}, in both disorder and interaction strength.  The plot shows that the transition to ETH in the model resembles  those obtained in other non-topological models (see e.g. \cite{Luitz2015}). However it also clearly illustrates that the measures of ETH and topological order cut through the parameter space in a very different orientation, showing that in a very literal sense that the phenomena are distinct and orthogonal.

Our second key result concerns competing ETH-violating effects in the model, which results in anomalous entanglement properties at high energies. Specifically we find that in certain regimes of parameter space, weak amounts disorder can lead to an {\em increase} in the averaged entanglement entropy. This property is observed in both the full- and  one-body entropies  (e.g. those calculated from the single particle density matrix \cite{Chung2001,Peschel03,Peschel04,Peschel09,Zhang2009,BrayAli2009}).  
We argue that this unusual behaviour is also related to effects  observed in the study of so-called many-body zero-modes \cite{Kells2018,Gangadharaiah2011,Goldstein2012,Yang2014,Kells2015a, Kells2015b,Fendley2016,Kemp2017,McGinley2017,Miao2017,Jermyn2014,Moran2017,Pellegrino2019,Mahyaeh2019}.  

This phenomena can be cast in the language of pre-thermalisation where the observed protection away from the ground state is due to an approximate conservation law \cite{Else2017}. Our results show that for entanglement measures of ETH/MBL  this effective conservation means that interactions can drive the system towards different types of thermal ensemble averages in agreement with the general predictions of \cite{Rigol2007,Rigol2008,Kollar2011,Rigol2011,Eisert2011,Vidmar2016}.  However we also show that disorder generally degrades this conservation law, and it is this effect that triggers the anomalous entanglement properties.

The structure of our paper is as follows. In section II we first review the Kitaev chain and relevant results with respect to both disorder and interactions, and address some practical difficulties associated with measuring the degree of topological order in small systems. In section III, we discuss the relevant MBL literature, focusing in particular on how the occupation-gap methodology can be extended to superconducting systems.  Some practical details are also provided on the calculation of one-body entanglement entropies. In section IV we discuss our main results in detail. We also provide several appendices where we explain some of the technical elements of our work in more detail and provide some additional evidence to back up our main claims. 

\begin{figure}
\centering
\includegraphics[width=0.35\textwidth]{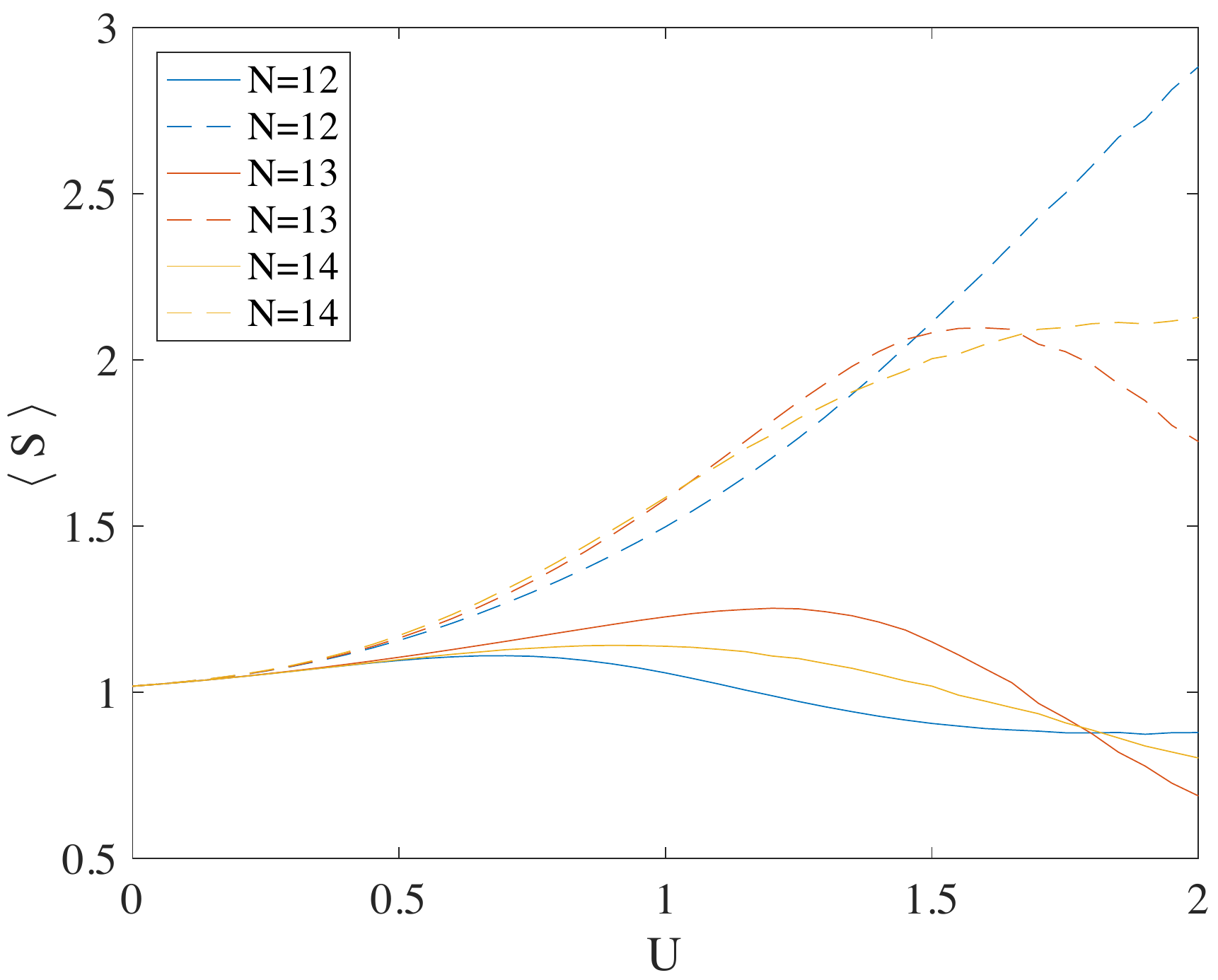}
\caption{ Entanglement entropies along line $W=0.1$ for even ground state for sizes $N=12,13,14$, $\mu=0.4,\, \Delta=0.7$, $500$ realisations. The one-body entanglement entropy (dashed) tends to be generically higher than the full entanglement entropy (solid).  The difference is minimal however for weakly interacting systems. }
\label{fig:Entropy_comparison_interaction_e0}
\end{figure}

\section{The Model and setup} 
\label{sect:Modelandmethods}
We formulate our results via the Kitaev chain model which is a lattice p-wave superconducting model \cite{Kitaev2001}:
\begin{align}\label{H:Kitaev}
H_0=-\sum_{j=1}^N \mu_j \left(c_{j}^\dag c_ {j}-\frac{1}{2}\right) - \sum_{j=1}^{N-1} t c_{j}^\dag c_{j+1} 
 + \Delta c_{j}^\dag c_{j+1}^\dag +{\rm h.c.}
\end{align}
where $t$ is the hopping parameter, $\Delta$ the pairing potential, and $\mu_j$ the local chemical potential at site $j$. Disorder is introduced by sampling each $\mu_j$ uniformly around a mean value $\mu$ with the deviation set by a parameter $W$. Namely $\mu_j$ are uniformly sampled from the interval $[-W+\mu,\mu+W]$. Interactions are introduced by local quartic term giving a density-density type interaction:

\begin{equation}
\label{H:Int}
H_{I} =2U \sum_{j=1}^{N-1} \left(c_j^\dag c_j -\frac{1}{2}\right) \left(c_{j+1}^\dag c_{j+1} -\frac{1}{2}\right). 
\end{equation}
With $U$ set to zero, the clean model is known to exhibit a topological phase when $|\Delta|>0$ and $|\mu|<2t$, with exponentially localised Majorana zeros modes at the ends of the wire \cite{Kitaev2001}. In the analysis $\Delta$ can be assumed real and in what follows $t$ will be set to one. The goal of our analysis is to ascertain whether the symmetry protected topological order in this model survives the transition from ETH behaviour (Eigenstate Thermalization Hypothesis \cite{Deutsch1991,Srednicki1994,Tasaki1998})  to what is known as MBL (Many-Body-Localization), see \cite{ETHMBL}.

\section{ Methods}
In this section we outline the methodologies we use to measure the transition from ETH to an MBL-type phase. Our main tool is a generalisation of the occupation gap method \cite{Bera2015,Bera2017} for superconducting systems that was used previously for this model in \cite{Kells2018}. This method has been shown to be a particularly sensitive signature of many-body-localization. Another advantage however is that it naturally allows us to probe so called one-body entanglement entropies, which allow us to find and quantify the traces of the topological order in both low and  higher energy states. 
 
\begin{figure*}
\centering
\includegraphics[width=0.99\textwidth]{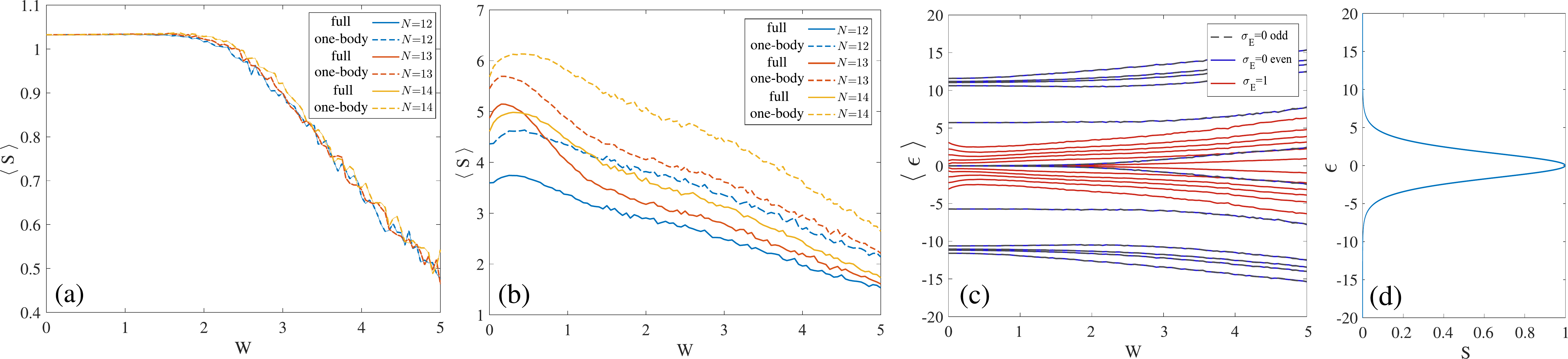}
\caption{Half-cut entanglement entropies along line $U=0.1$ for system sizes $N=12,13,14$, $\mu=0.4, \Delta=0.7$,  (a) Ground state $\sigma_E=0$ - for the ground states of a weakly interacting system there is minimal difference between the averaged full and one-body entanglement entropies. (b)  Averaged $\sigma_E=1$ one-body entropy is consistently higher than the full entanglement entropy - the same general trends are observed however  (c) Each mode in the quasi-particle entanglement spectrum contributes to the full entanglement entropy according to  (d) [see \eqref{eq:Se}] }
\label{fig:Entropy_explain}
\end{figure*}

\subsection{Generalised Single particle density matrix and the occupation gap}
In (\cite{Bera2015,Bera2017}) it was argued that the discontinuity in the single particle occupation numbers that occurs in non-interacting systems should also persist if the system is in the MBL phase. On the other hand this occupation gap would  be completely washed out in an ergodic (ETH) phase, forming a smooth function in the thermodynamic limit.  In order to study the region of overlap between an MBL-type phase and the topological phase of the model we use a generalization of this method to superconducting systems (see e.g. \cite{Kells2018} and references therein).  In this method one first forms the generalised single particle density matrix (GSPD herein):
\begin{equation}
\label{R:mat}
\mathscr{R} =\begin{pmatrix}\rho & \kappa \\ -\kappa^* & 1-\rho^*\end{pmatrix}, 
\end{equation}
where $\rho_{ij}=\expval{c^\dagger_j c_i}{\Phi}$ and $\kappa_{ij}=\expval{c_j c_i}{\Phi}$. (For properties of this matrix see Appendix \ref{App:GSPD}.)  For the analysis of the single particle entanglement we will also need to define a reduced density matrix $\mathscr{\tilde{R}}$ where we only consider a subset of matrix indices, corresponding to a partition of the sites in position space.

Given an eigenstate of the Hamiltonian one can construct the GSPD corresponding to this state, which upon diagonalization, yields so called natural single particle orbitals \cite{Bera2015}.
\[\mathscr{R}\ket{\phi_\alpha}=n_\alpha\ket{\phi_\alpha}\]
In the context of superconductivity or more generally particle non-conserving Hamiltonians, the single particle orbitals will be Bogoliubov quasi-particle orbitals, with eigenvalues $n_\alpha$ are interpreted as occupations with $0\leq n_\alpha\leq 1$.

In the non-interacting limit the numbers $\Delta n = n_{N+1} - n_{N}$ display a sharp jump from zero to one, delineating empty and filled orbitals. This jump gradually becomes less pronounced as we increase the local interaction strength and move to higher energy densities.  In this case introducing disorder tends to on average reduce this trend, reinstating a sharper occupation gap.  This gap then offers a means to distinguish ergodic $\Delta n \sim O(1/N) $ from localised $\Delta n \sim O(1)$ phases. 

\subsection{Full and One-Body entanglement entropies}
To further probe the region of overlap between MBL-type and topological phases of the model, we look at the full- and one-body entanglement entropies, performing position space cuts near the centre of the wire, of both ground ($\sigma_E = 0$) and high energy states ($\sigma_E = 1$). The energy density $\sigma_E$ (denoted $\epsilon$ in \cite{Bera2017}) is defined as $\sigma_E=2(E-E_{\text{min}})/(E_{\text{max}}-E_{\text{min}})$, for a given disorder realisation.  

One expects, due to the spatial localisation of the many-body wavefunctions, that the entanglement entropy in an MBL phase to be generally lower than in the ergodic phase. Moreover it can be shown that  one of the defining features of the MBL phase is an area law scaling for entanglement entropy which, for one-dimensional systems, should be constant, or at least bounded by a constant \cite{Bauer2013,Nandkishore2015}. The (generalized) one-body approximation to the entanglement entropy  is calculated directly from the reduced (generalized) single particle density matrix. For more details see appendix \ref{App:GSPD}. In the case of free fermion systems this calculation agrees exactly with the standard entanglement entropy (see \cite{Chung2001,Peschel03,Peschel04,Peschel09,Zhang2009,BrayAli2009}).  As one moves to more interacting regimes the one body entropy generically overestimates the full entropy (see Figures  \ref{fig:Entropy_comparison_interaction_e0}, and \ref{fig:Entropy_explain}), although for modest interaction strengths the results are very similar.

Our main motivation for examining the one-body entropy is that it allows us to break down the total entanglement entropy into the contributions from the individual modes of the so-called entanglement Hamiltonian. Concretely in this case we have
\be
S=-\sum_{\varepsilon_i>0} f(\varepsilon_i) \log_2 (f(\varepsilon_i))+ f(-\varepsilon_i) \log_2 (f(-\varepsilon_i))
\label{eq:Se}
\ee
where $f(\varepsilon) = (1+ e^{\varepsilon})^{-1}$ and the values $f(\varepsilon_i)$ are the eigenvalues of the reduced generalised density matrix $\tilde{\mathscr{R}}$ where the subset are the position space indices for half of the wire (e.g. $j \le N/2$). The values $\varepsilon_i$ themselves are known as the quasi-particle entanglement spectrum \cite{BrayAli2009}, which generate the entanglement spectrum.

\subsection{Thermal and Generalised Gibbs Ensembles}
To study the expected MBL and ETH phases of the model we compare numerical results to  both the thermal (grand canonical) and  a generalised Gibbs ensemble defined as:
\begin{align}
\rho_{\text{th}}&=\frac{\exp(-\beta H)}{\mathcal{Z}_{\text{th}}},\\
\rho_{\text{G}}&=\frac{\exp(-\beta H-\alpha \mathcal{N})}{\mathcal{Z}_{\text{G}}},
\end{align}
where, for simplicity,  $H$ represents the clean ($W=0$) Kitaev-Hubbard system,  and $\mathcal{N}$ is the quasi-particle number operator for the clean non-interacting ($W=U=0$) model  but where we exclude the zero mode number operator. For definition and discussion see \cite{Nqop}.  Here $\alpha$ plays the role of quasi particle chemical potential \cite{Rigol2007,Rigol2008,Kollar2011,Rigol2011,Eisert2011,Vidmar2016}. In the thermodynamic limit \cite{Nandkishore2015} one expects that if ETH holds then for a small subsystem $A$ with environment $B$, the reduced density matrix of an eigenstate will agree with the reduced density matrix of $\rho_{\text{th},A}=\tr_B(\rho_{\text{th}})$, with $\beta$ fixed to give the expected energy of the eigenstate. In particular, this means that the von Neumann entropy of $\rho_{\text{th},A}$ will agree with the entanglement entropy of the eigenstate.

To examine the global trends of entanglement entropy of eigenstates, we study the von Neumann entropy of the reduced thermal matrix $\rho_{\text{th},A}$ as a function of inverse temperature ($\beta$) \cite{ThermalEE}. For a fixed chain length, the infinite temperature limit $\beta=0$, corresponds to $\rho_{\text{th}}=\Bbb{I}/2^N$, the normalised identity operator, which gives $E_{\text{avg}}=\tr(H)/2^N$ for the expected energy, and the von Neumann entropy of the reduced density matrix $\rho_{\text{th},A}$ peaks at $\log_2(D_A)$ where $D_A$ is dimension of reduced Hilbert space of subsystem $A$. If we allow for negative temperatures we can compare this to the entanglement entropy of the eigenstates across all energies $E$, where $E<E_{\text{avg}}$ corresponds to positive temperatures, and $E>E_{\text{avg}}$ corresponds to negative temperatures. 

In order to describe trends in entropy-energy of eigenstates, which in the weakly interacting chain still have a near integer expected number of quasi-particles, we also compare against the von Neumann entropy of the reduced Gibbs-type density matrix $\rho_{\text{G},A}$ as a function of inverse temperature.  We impose the constraint $n=\tr(\mathcal{N}\rho_{\text{G}})$, with $n$ an integer $0\leq n\leq N-1$. The Gibbs-type operator is the one which maximizes the statistical entropy given this constraint, which is normally arrived at by treating $\alpha$ as a Lagrange multiplier. The trace condition fixes $\alpha(\beta,n)$ as a function of $\beta$, and quasi-particle number $n$. Numerically however we would like to find such an $\alpha$ to construct the density matrix, so for fixed values $\beta,\,n$ we employ a bisection method on the parameter $\alpha$.

\section{Numerical Results}

Our key goal is to ascertain if Fock-space localisation and topological superconductivity can coexist in this model. To analyse this in Figure \ref{fig:phases14} we overlay contour plots of the values of $\langle\Delta n\rangle$ (in blue) and the inverse effective coherence length $\langle 1/\xi\rangle$ (in red) (see appendix \ref{App:Top}), for a system size $N=14$.  The edge of the topological region is in principle  where $\langle 1/\xi\rangle \to 0$. In practice however, non-exponential decay could be indicated by values of $\langle 1/\xi\rangle  \to O(1/N)$.  Similarly for a ETH-MBL boundary in the thermodynamic limit  one expects that the occupation gap $\langle\Delta n\rangle$  in ergodic regions to be zero, such that  any non-zero-value would signal the MBL phase. For finite systems however one always expects some small discontinuity and so more realistically the values of $O(1/N)$ should be taken as an indication of the threshold. 

In Figure \ref{fig:phases14} we see that around the non-interacting line (x-axis) there is a range of parameters where the measures for topological and Fock-space localisation coexist.  One interesting feature is that around these parameters ($\mu = 0.4$ and $\Delta=0.7$)  there is a region which maintains a significant occupation gap even for perfectly clean systems. We will argue below that this is related to a peculiarity in the band structure of this model, which leads to an approximate conservation law that offers some topological protection of high-energy states in the presence of weak interactions \cite{Else2017}.  The effect reduces when bands of excited states with different quasi-particle number merge more completely.  This occurs for example when one moves to weaker p-wave pairing and/or value of chemical potential that are closer to the bottom of the conduction band.  However, as discussed in more detail below,  disorder also breaks this approximate conservation. We show below that this explains why introducing small amounts of disorder into the model can actually lead to an increase in the average entanglement entropy. 

\begin{figure}
\centering
\includegraphics[width=0.5\textwidth]{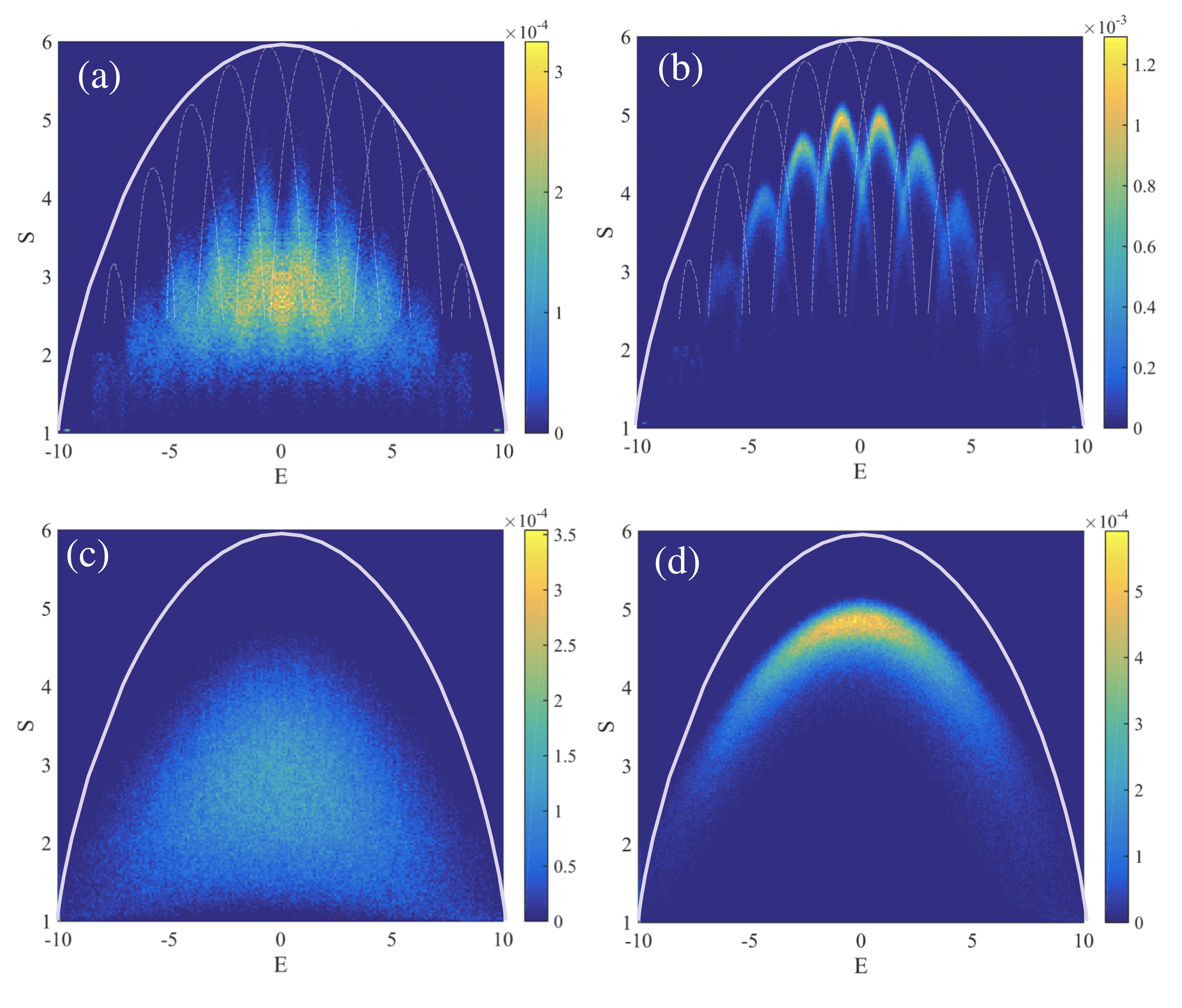}
\caption{ $N=12$ sites. Entanglement entropy versus Energy for $\mu=0.4$ and $\Delta=0.7$ for (a) $U=0.1$, $W=1$ (b) $U=0.2$, $W=0.3$ (c) $U=0.1$, $W=2$ (d) $U=0.5$ and $W=2$.}
\label{fig:EntropyEnergy}
\end{figure}

\subsection{Entanglement structure of coexisting region}
 In order to further examine the interplay between ETH-violating mechanisms and topological superconductivity we examine how the entanglement entropy of a bi-partition is changed as we increase both the interaction strength $U$ and the disorder parameter $W$, focusing on both ground-state ($\sigma_E =0$) and excited states ($\sigma_E=1)$. The ground state properties are shown in Figures \ref{fig:Entropy_comparison_interaction_e0} and \ref{fig:Entropy_explain} (a) where the full-  and one-body- entanglement entropy for systems sizes $N=12,13,14$ are calculated in the even parity sector and disorder averaged for $W=0.1$ (constant $W$) and $U=0.1$ (constant $U$) lines respectively.
  
For the ground state properties we see that for a significant region of the disorder parameter space the entanglement entropy remains roughly constant ($\sim 1$ entangled bit), before eventually becoming reduced at around $W=2$.  The lack of scaling is expected for the ground state of a gapped system following an area law in entanglement entropy. In Figure \ref{fig:Entropy_explain} (c) we also plot the quasi-particle entanglement spectrum for the $N=12$ system,  from where we can see how the individual one-body quasi-particle entanglement spectrum values contribute to the total.   From Figure \ref{fig:Entropy_explain} (c) (blue and dashed grey data) we see that the one-body contribution that dominates is a pair of almost zero eigenvalues, which slowly start to gap out as the system becomes more disordered.  
  
For excited states the results are significantly different.  Firstly in  \ref{fig:Entropy_explain} (b) we see that the entropy values in low disorder regions are significantly larger than the ground state measures. However, we also see that for these parameter values, modest amounts of disorder actually increase  entropy even further before eventually causing it to reduce.   The fingerprints of this behaviour can also be seen clearly in the quasi-particle entanglement spectrum in Figure \ref{fig:Entropy_explain} (c) (red data).  There we see that the generally higher entropy is caused by an accumulation of the quasi-particle spectrum around the $\varepsilon=0$ point, where the states contribute the most weight [see e.g. Figure \ref{fig:Entropy_explain} (d), and equation \eqref{eq:Se}].  The initial increase in total entropy can be understood as a tightening of this entanglement spectrum band around the $\varepsilon=0$ line, before eventually dispersing at larger $W$ values.  In the next section we will show that this behaviour can be understood by examining the relationship between entropy and the many-body spectrum.
 
\subsection{Entanglement -v- energy spectrum  - ensemble switching} 
\begin{figure}
\centering
\includegraphics[width=0.4\textwidth]{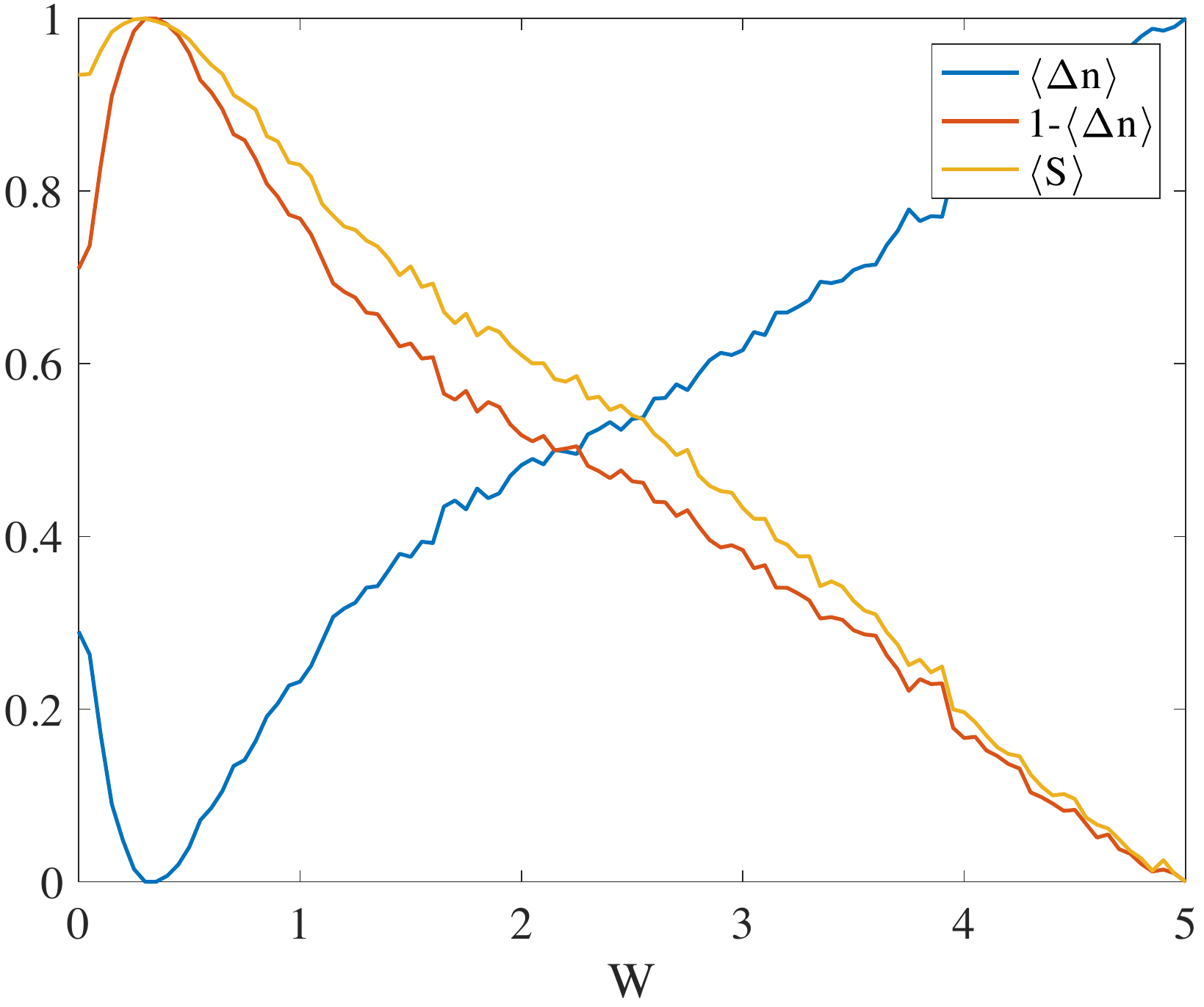}
\caption{ Normalized entanglement entropy, and occupation gap with increasing disorder strength, for $N=12$, $\mu=0.4$, $\Delta=0.7$ and $U=0.1$, averaged for $1000$ disorder realisations. The normalisation is a shift followed by a rescale $x\to (x-x_{min})/(x_{max}-x_{min})$, where, $x=\langle S\rangle,\langle \Delta n \rangle$ and $x_{max/min}$ are the max/min values of the averaged quantity along the line $U=0.1$ with $0\leq W\leq 5$ as shown in the figure. The negative correlation can be seen in a density plot for all eigenstates of the model for $100$ disorder realisations in Figure \ref{fig:EntropyvsDn}.
 }
\label{fig:NormEntropyDn}
\end{figure}
While the expected trend is that disorder should decrease the average value of the entanglement entropy, in Figure \ref{fig:Entropy_explain} we first see an initial increase  before the onset of the expected decrease. Similar anomalous behaviour can be seen in the lower left corner of Figure \ref{fig:phases14} where we see a dip in the blue contours indicating drop in $\langle\Delta n\rangle$ with disorder. As noted in \cite{Bera2017}, there is a strong negative correlation between $\Delta n$ for individual eigenstates and their entanglement entropies. Indeed it was shown that in the ETH phase, eigenstates tend to have large values of entanglement entropy and near-zero values of the occupation gap. In contrast after the transition to the many-body-localized regime, the occupation gap tends to unity, while the eigenstates have comparatively low values of entropy. The same correlation is shown in Figure \ref{fig:NormEntropyDn}, where the averaged entanglement entropy for $\sigma_E=1$ states has been normalised to lie within the same range as $\Delta n$, and is overlaid with rescaled values of $\langle \Delta n \rangle$ (and $1-\langle \Delta n \rangle$ to show the direct correlation). 

In order to pose an explanation for the anomalous increase in entanglement entropy with moderate disorder strength,  we study the global properties of entanglement entropy of all eigenstates. The results are plotted in Figure  \ref{fig:EntropyEnergy}. We identify four distinct behaviours which reflect the interplay between interaction and disorder strength.  In particular, we identify regions of parameter space where  single particle bands weakly overlap corresponding to relatively weak interaction strength. Here,   the energy-entropy plots arrange into a pattern of multiple overlapping `parabolas'  as  shown in Fig.  \ref{fig:EntropyEnergy} (b). This behaviour is  consistent the von Neumann  entropy $S$ of the reduced density matrices of the Gibbs type $\rho_{\text{G}}$  plotted against the expected value of the energy $E=\tr(H\cdot \rho_{\text{th/G}})$ as the inverse temperature $\beta$ is varied (dashed line), and reflects an approximate particle number  conservation law in this regime. 

Strong interactions and moderate disorder break this approximate conservation law  resulting in a global parabolic trend  Fig.  \ref{fig:EntropyEnergy} (d). This  is consistent the von Neumann  entropy $S$  of the thermal ensemble $\rho_{\text{th}}$ (solid line) and appears in regions of parameter space with negligible occupation number discontinuity (see Fig. \ref{fig:phases14}). Disorder scatters these global  structures leading to a reduction in the entanglement  entropy which  deviates from the thermal or Gibbs ensemble,  reflecting the violation of ETH seen in Fig.   \ref{fig:EntropyEnergy} (a) and (c).

The anomalous increase in entanglement entropy occurs when crossing from regimes where the behaviour is of Gibbs type  \ref{fig:EntropyEnergy} (b) to regimes approaching the thermal type  \ref{fig:EntropyEnergy} (d). This behaviour is then cutoff by a decrease in entanglement entropy caused  by increased disorder \ref{fig:EntropyEnergy} (c).}
We note that  regions of parameters space characterized by  bands that weakly overlap (as in Fig.  \ref{fig:EntropyEnergy} (b)),  and thus reflect an approximate particle number conservation,  have been shown to protect  many body zero modes \cite{Kells2015a, Kells2015b, Kemp2017,Else2017}.  However, here moderate disorder tends to drive the system to regimes where stronger band overlaps occur. This is expected to slightly increase the entanglement entropy and was shown to  reduce the topological protection of the many body zero modes \cite{Kells2018}. Stronger disorder then localizes the single particle states and consequently  cuts off  many body effects. This is expected to reduce the entanglement entropy and partially restore  the protection of the zero modes before eventually driving the system entirely out of the topological phase. 

Here we stress that while this anomalous effect is observed for small values of $U$ and $W$, where one would expect to observe signatures of the topological phase, the effect itself seems unlikely to be purely topological in nature. For instance, in Figures \ref{fig:n13entdn} and \ref{fig:n13entdnpb}, for $N=13$ sites, and $\mu=-3$, $\Delta=0.7$, with the clean non-interacting model outside of the topological phase, we see the same increase in entanglement entropy and decrease in $\Delta n$ for small disorder and weak interactions.  In addition previous studies of global properties of the entanglement entropy in models such as an XXZ ladder, Bose Hubbard model \cite{Haque2015}, have shown  the same global versus multi-parabolic trends when plotting entanglement entropy vs energy.

{\em Conclusion}:
We analysed the behaviour of the disordered Kitaev-Hubbard model with the goal of identifying   the coexistence of localization  and symmetry protected topological order. Our methodology uses  eigenstate entanglement entropy and  a generalization of the single particle occupation gap for superconducting systems, and 
 demonstrates  that similar to  its number conserving counterpart \cite{Bera2017},  there is a strong correlation  between the averaged value of entanglement entropy and the generalized occupation gap for $\sigma_E=1$ states.

Our analysis allows us to discern distinct mechanisms leading to  ETH violation, which can  be attributed to pre-thermalization effects or disorder. In particular, our results identify  a narrow region of parameter where SPTO and  disorder induced ETH violation can co-exist. 

In our study we discovered an unexpected increase of the entanglement entropy due  to moderate amounts of disorder. This phenomena occurs when the interacting system transitions between  different types of thermal ensembles: The relatively clean system  follows  a generalized Gibbs-like ensemble with an approximate number conservation law (also known as pre-thermalization).  Moderate disorder shifts the system slightly towards a thermal ensemble, which is eventually cutoff by strong disorder, which in turn reduces the entanglement  entropy compared to the thermal ensemble.  

\newpage

{\em Acknowledgments}
We are grateful for stimulating conversations with Aaron Conlon, Domenico Pellegrino, Luuk Coopmans, Joost Slingerland,  Paul Watts, Shane Dooley, Jiannis Pachos, Fabian Heidrich-Meisner, Masud Haque and Jens Bardarson.  S.N. acknowledges the Government of Ireland Postgraduate Scholarship GOIPG/2014/150 provided by the Irish Research Council. J.V. was funded in part by Science Foundation Ireland under the Principal Investigator Award 10/IN.1/I3013. D.M. acknowledges support from the Israel Science Foundation (Grant No. 1884/18). G. K.  was supported by a Schr\"{o}dinger Fellowship and acknowledges support from Science Foundation Ireland through Career Development Award 15/CDA/3240.

\appendix
\section{Generalised Single Particle Density Matrix}
\label{App:GSPD}
Given a fixed number $L$ of fermionic degrees of freedom, and many body fermionic state $\psi$, not necessarily an eigenstate of the total number operator $\hat{N}$, we can form the generalised single particle density matrix $\mathscr{R}_{\psi}$ by calculating the expectation values of quadratic fermion operators $c_i$ and $c^\dagger_j$, $1\leq i,j\leq L$. That is defining the following matrices,

\begin{align*}
\rho_{ij}&=\expval{c^\dagger_j c_i}{\psi}\\
\kappa_{ij}&=\expval{c_j c_i}{\psi}\\
\mathscr{R}_\psi&=\pmqty{\rho & \kappa\\ -\kappa^* & 1-\rho^*}
\end{align*}
This matrix in block form above has the property that it can be used to calculate expectation values in the state $\psi$ of all operators quadratic in the creation and annihilation operators. Writing an operator $\hat{M}=\sum_{i,j}A_{ij}c^\dagger_i c_j+B_{ij}c^\dagger_i c^\dagger_j+C_{ij}c_i c_j+D_{ij}c_i c^\dagger_j$ in matrix form by collecting the fermi  operators in vector form,

$$\hat{M}=\pmqty{c^\dagger & c}\pmqty{A & B\\ C & D}\pmqty{c \\ c^\dagger}, \quad M=\pmqty{A & B\\ C & D}$$
then we have that $\expval{\hat{M}}{\psi}=\tr(\mathscr{R}_\psi M)$. 

Suppose we perform a Bogoliubov transformation of the fermionic operators, $\beta^\dagger_i=\sum_j U_{ji} c^\dagger_j +V_{ji} c_j $, requiring the transformation to be canonical and that $\beta_i=(\beta^\dagger_i)^\dagger$, we find that the matrix $W=\pmqty{U & V^* \\ V & U^*}$ is unitary, and the transformation can be written as $\pmqty{\beta \\ \beta^\dagger}=W^\dagger\pmqty{c \\ c^\dagger}$. If we form the GSPD matrix of the quasi-particle vacuum state of this transformation, that is the state $\ket{\phi}=\prod \beta_i\ket{0}$, then matrix can be calculated as $\mathscr{R}_\phi=\pmqty{V^*\\U^*}\pmqty{V^T & U^T}$. In changing to the quasi-particle basis, GSPD matrix of the vacuum state can be put in its canonical form 
$$\mathscr{R}_\phi'=W^\dagger \mathscr{R}_\phi W=\pmqty{0 & 0 \\ 0 & 1}=\pmqty{\expval{\beta^\dagger \beta}{\phi} & \expval{\beta \beta}{\phi} \\ \expval{\beta^\dagger \beta^\dagger}{\phi} & \expval{\beta \beta^\dagger}{\phi}},$$
from which it is clear that the matrix satisfies $\mathscr{R}_\phi^2=\mathscr{R}_\phi$.

\begin{figure}[h!]
\centering
\includegraphics[width=0.4\textwidth]{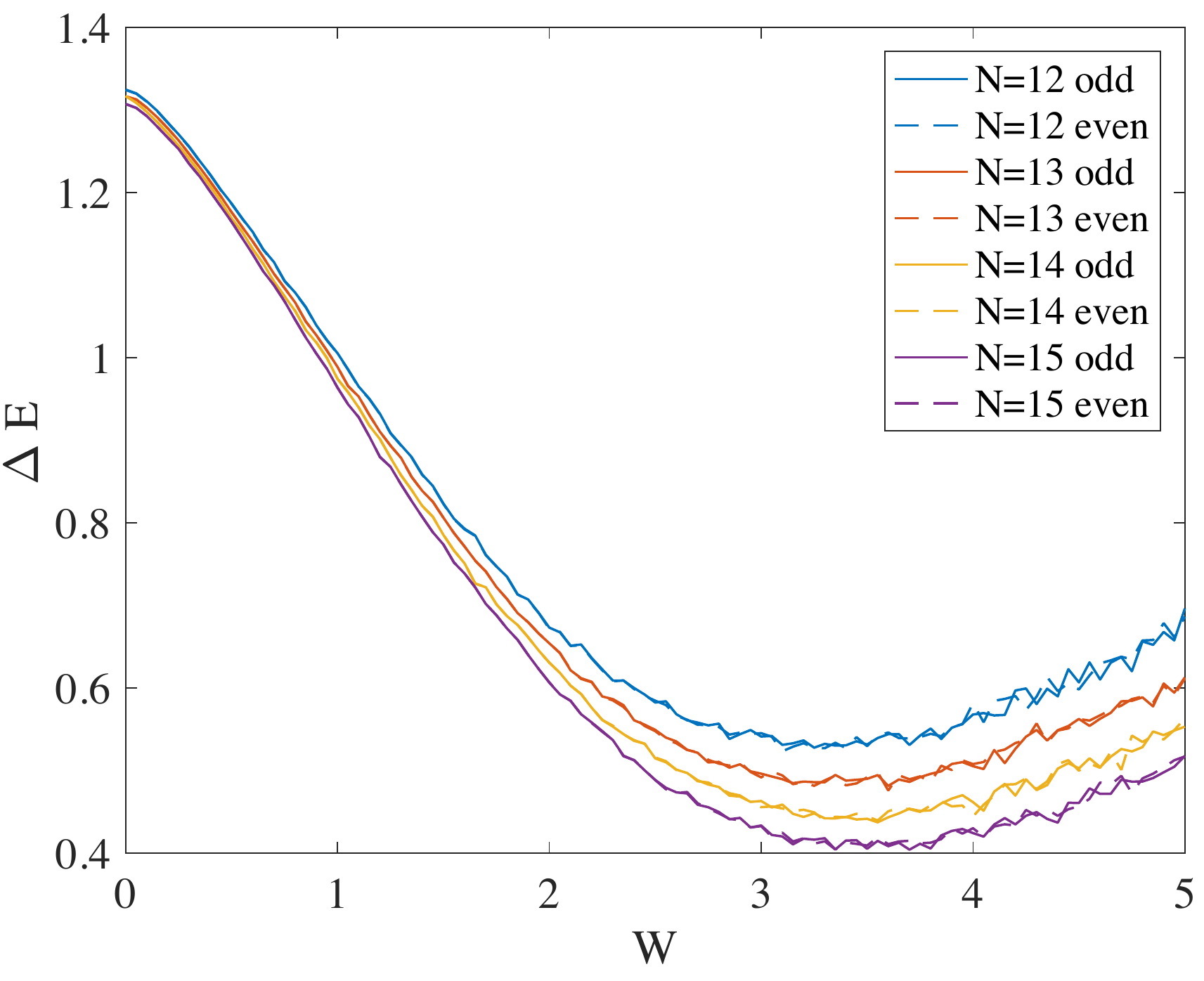}
\caption{\,Energy gap for both number parity sectors as a function of disorder strength with fixed interaction strength $U=0.1$, $\mu=0.4, \Delta=0.7$ with $2000$ disorder realisations}
\label{fig:oddgapcloses2}
\end{figure}

\begin{figure}[h!]
\centering
\includegraphics[width=0.4\textwidth]{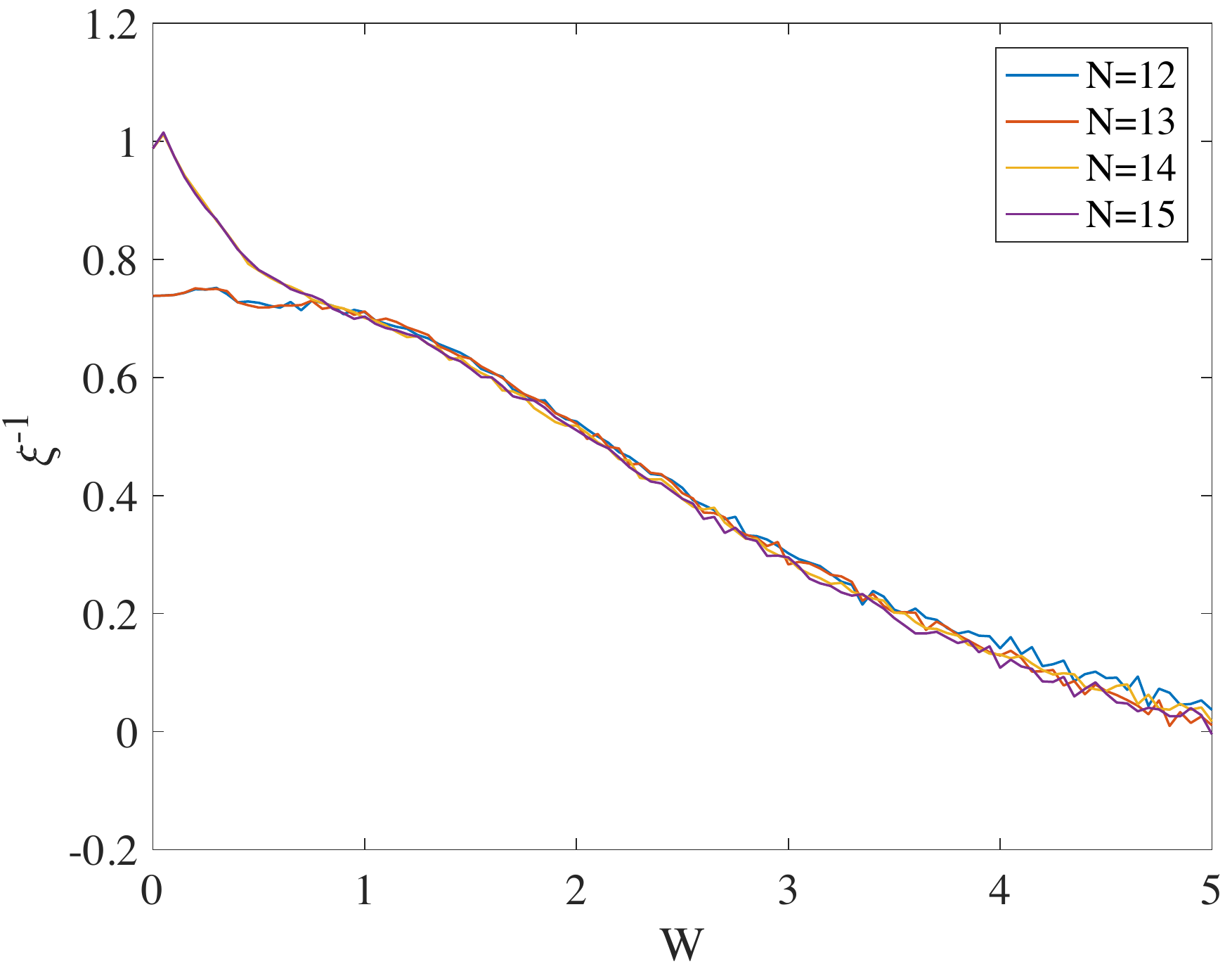}
\caption{\,Inverse coherence length as a function of disorder strength and fixed interaction strength $U=0.1$, $\mu=0.4, \Delta=0.7$ with $2000$ disorder realisations}
\label{fig:loclengthvsW}
\end{figure}

\subsection*{One Body Entanglement Entropy}
\begin{figure}
\centering
\includegraphics[width=0.4\textwidth]{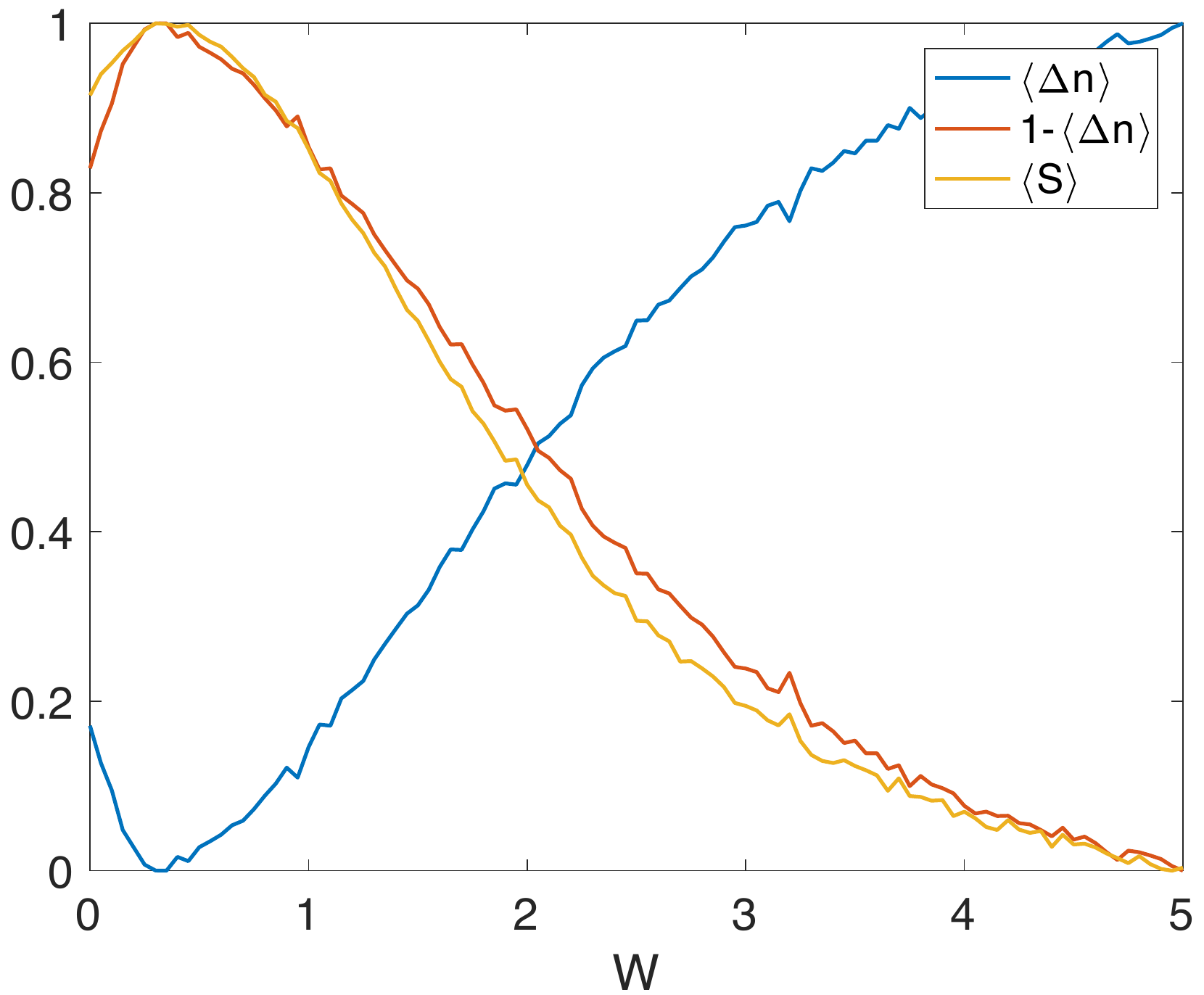}
\caption{$N=13$ sites, $\mu=-3$, $\Delta=0.7$, $U=0.1$, $500$ disorder realisations. (a)Scaled entanglement entropy and $\Delta n$ vs disorder strength.}
\label{fig:n13entdn}
\end{figure}

\begin{figure}
\centering
\includegraphics[width=0.4\textwidth]{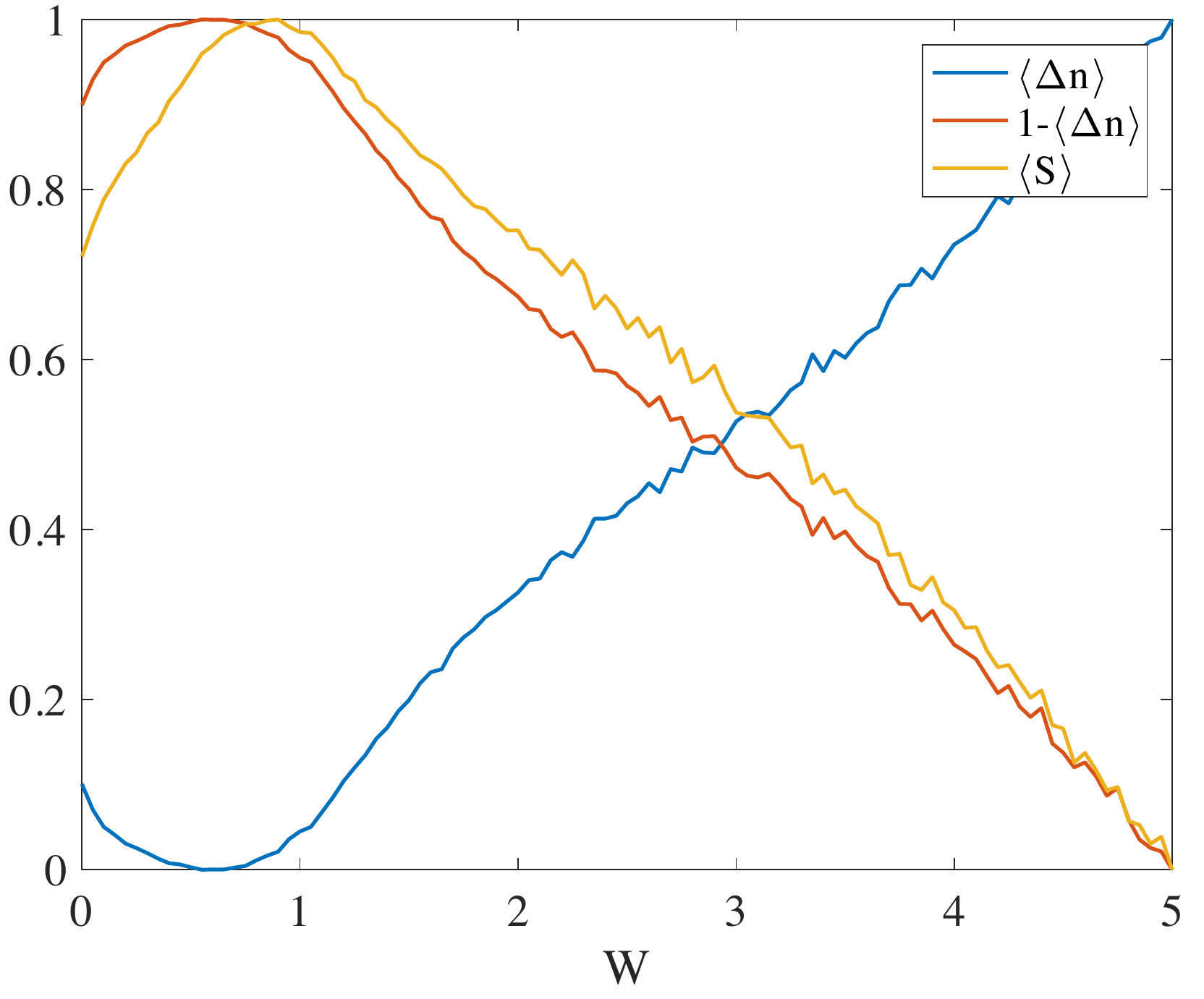}
\caption{$N=13$ sites, $\mu=-0.2$, $\Delta=0.7$, $U=0.1$, $500$ disorder realisations. (a)Scaled entanglement entropy and $\Delta n$ vs disorder strength, periodic boundary conditions.}
\label{fig:n13entdnpb}
\end{figure}

Choosing a subset of the labels (sites) of the fermion operators $\{j_1\ldots j_k\}$ we can form a reduced density matrix $\widetilde{R_\psi}$ by taking only the matrix entries of $R_\psi$ corresponding to the indices $\{j_1\ldots j_k\}$, which are the expectation values $\expval{c^\dagger_{j_l} c_{j_m}}{\psi}$, with $1\leq l,m\leq k$. This is equivalent to deleting rows and columns for labels (sites) that are to be traced out.

Now if one takes the von Neumann entropy of this reduced density matrix, we call this the one body entanglement entropy $S_{\text{one}}=-\tr(\widetilde{R_\psi}\cdot \log_2(\widetilde{R_\psi}))$. It was shown by \cite{Chung2001} that this is precisely the entanglement entropy provided that the state is a free fermion eigenstates. In this case we can write $\tilde{\mathscr{R}}=(1+\exp(H_\alpha))^{-1}$ \cite{Blaizot1986}, where $H_\alpha$ is free fermion single particle/excitation Hamiltonian. Given that the eigenvalues of $\tilde{\mathscr{R}}$ come in pairs $\frac{1}{2}\pm\eta_i$ due to particle hole symmetry, this means the eigenvalues of $H_\alpha$ come in pairs $\pm \epsilon_i$. The numbers $\epsilon_i$, are called the quasi-particle entanglement spectrum \cite{BrayAli2009}. The entanglement entropy computed from the quasi particle entanglement spectrum is $-\sum_i f(\epsilon_i)\log_2(f(\epsilon_i)) $, where $f(\epsilon)=(1+e^{\epsilon})^{-1}$\cite{BrayAli2009}. While this function isn't symmetric with respect to $\epsilon$, given that the quasi-particle entanglement spectrum is particle hole symmetric, we can combine the positive and negative pairs into a symmetric contribution which peaks at $\epsilon=0$.

For a generic state with a definite fermion number parity the one-particle entropy tends to overestimate the entanglement entropy.  This is clearly seen in particular in Figure \ref{fig:Entropy_explain} where even for small interaction strengths the average value for $\sigma_E=1$ states is approximately one e-bit above the actual value of the entanglement entropy. That said, it is also clear that for the $\sigma_E=0$ states, even for modest interactions that the value of the one body entropy reasonably accurately estimates the actual value of the entanglement entropy. Furthermore, the one-body entropy tends to mimic the global trends in in entropy seen in Figure \ref{fig:EntropyEnergy}.

\section{Interaction and Topological Phase transitions for finite systems}
\label{App:Top}

\begin{figure}
\centering
\includegraphics[width=0.45\textwidth]{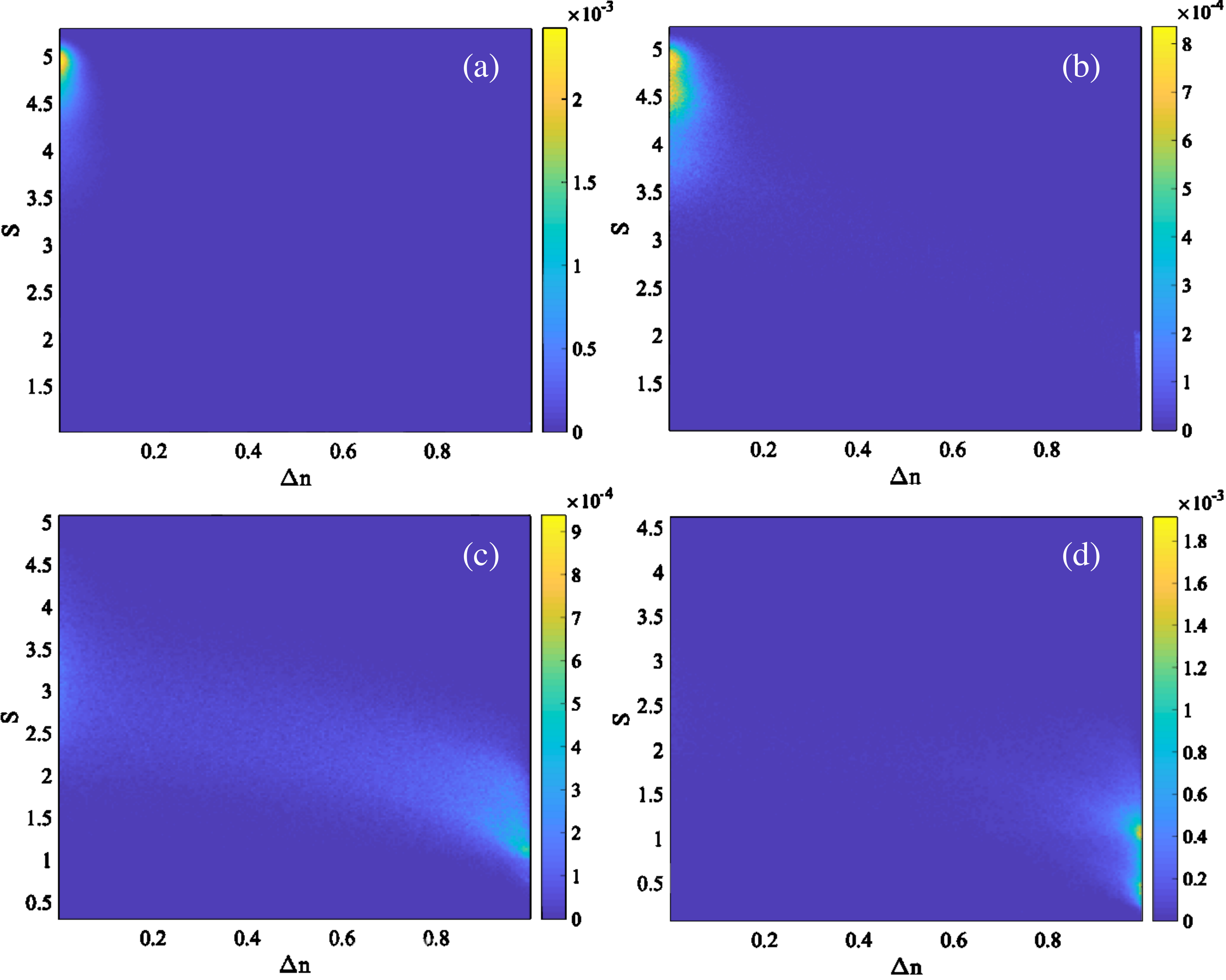}
\caption{$N=12$ sites. Entanglement entropy versus occupation gap for $\mu=0.4$ and $\Delta=0.7$ for (a) $U=0.5$, $W=0.5$ (b) $U=0.2$, $W=0.3$ (c) $U=0.1$, $W=3$ (d) $U=0.1$ and $W=5$}
\label{fig:EntropyvsDn}
\end{figure}

It is known that topological superconductivity survives for a range of disorder and interaction parameters \cite{Motrunich2001,BDRvOProb2011,BDRvOTop2011,Akhmerov2011,RKDM2012,RBA2013,DeGottardi2013,Pientka2013,Stoudenmire2011,Sela2011,Lutchyn2011, Lobos2012,Crepin2014,Hassler2012,Thomale2013,Katsura2015,Gergs2016}.  In order to define the topological phase of the model one indicator is the gap. Another distinguishing feature is the presence of Majorana modes localised at the ends of the one dimensional wire. Addressing the first, in Figure \ref{fig:oddgapcloses2} the disorder averaged energy gap to the first excited state in the even/odd parity sector is plotted against disorder and interaction strength. 

A clear energy gap can be seen in this figure, and it is expected that the gap closes as interaction strength and/or disorder is increased. It is unlikely that zero will actually be obtained from this disorder averaging process given that the energy gap is calculated as a positive quantity, and averaging at best can yield a small but non-zero number. In order to show the expected closing of the energy gap, in Figure \ref{fig:oddgapcloses2} the even/odd parity energy gap is plotted for increasing system size for a fixed interaction strength $U=0.1$ which is averaged over $500$ disorder realisations. One can see the downward trend of the value of the gap at larger disorder and increasing system size.

In the main text we instead use the inverse of the effective Majorana coherence length $1/\xi$ for the zero modes in the topological phase of the model. Given the (even and odd parity ground states) ground state and first excited state, denoted $\ket{0}$ and $\ket{1}$ of the above model, we define the cross-correlators 
\begin{equation}
\bra{0}c_j\ket{1}+\bra{0}c^\dagger_j\ket{1}=\bra{0}\gamma_j\ket{1}\text{ for } j=1,\ldots,N.
\label{cross-cor}
\end{equation}
This function of position ($j$) decays exponentially into the middle of the wire as $e^{-x/\xi}$ where $x=j$ (or $N-j$ if using a minus sign in definition above). The effective inverse coherence length $1/\xi$ is estimated by taking the logarithm of absolute value of the cross correlators, averaging over disorder realisations and then performing a linear fit (least squares) from sites $1$ up to $\lfloor N/2 \rfloor$. Performing the linear fits first and then averaging over the disorder realisations gives an identical answer, which is a property of linear fitting with the least squares method.

The inverse coherence length is shown in Figure \ref{fig:loclengthvsW}, averaged over $2000$ disorder realisations for parameters $U/t=0.1, \mu/t=0.4$ and $\Delta/t=0.7$. The coherence length decays to zero relatively consistently when the disorder strength reaches approximately $W\sim 5$. The sudden jump in coherence length from sizes $12,13$ to $14,15$, is an artificial artefact of the procedure to estimate the coherence length.


\begin{thebibliography}{}
\bibitem{Nqop} To elaborate on the definition of the $\mathcal{N}$ operator, we take the clean ($W=0$) Kitaev model, diagonalising as usual with a Bogoliubov transformation so that $H_0=\sum_j E_j \beta^\dagger_j \beta_j +\Lambda$, $\Lambda$ being the ground state energy. In the topological phase there are (approximate) zero modes which are combined into a fermion degree of freedom with single particle energy $E_k\sim 0$ for some index $k$, then $\mathcal{N}$ is simply 
\begin{equation}
\mathcal{N}=\sum_{j\neq k} \beta^\dagger_j \beta_j
\end{equation}
With weak disorder and interactions, the energy spectrum maintains the approximate doubling of degeneracy with number parity. Inclusion of the zero mode number operator does not allow the assignment of a single number of quasi-particle excitations to a single band as in Figure \ref{fig:EntropyEnergy} (b), since occupying the zero mode d.o.f. counts as a quasi-particle with approximately the same energy. If one includes the zero mode, one sees curves as in \ref{fig:EntropyEnergy} (b) (dashed) which instead encompass two neighbouring 'parabolas' instead of just one.
\end{thebibliography}
\end{document}